# Antiferromagnetic α-MnTe: Molten-Salt-Assisted Chemical Vapor Deposition Growth and Magneto-Transport Properties

*Shuaixing Li,*[1,†] *Jianghua Wu,*[2,†] *Binxi Liang,*[1] *Luhao Liu,*[1] *Wei Zhang,*[3] *Nasrullah Wazir,*[2] *Jian Zhou,*[1] *Yuwei Liu,*[2] *Yuefeng Nie,*[2] *Yufeng Hao,*[2] *Peng Wang,*[2] *Lin Wang,*[3] *Yi Shi,*[1,*] *Songlin Li*[1,*]

[1]National Laboratory of Solid-State Microstructures, Collaborative Innovation Center of Advanced Microstructures, and School of Electronic Science and Engineering, Nanjing University, Nanjing University, Nanjing 210093, China
[2]College of Engineering and Applied Sciences and Jiangsu Key Laboratory of Artificial Functional Materials, Nanjing University, Nanjing 210023, China
[3]Key Laboratory of Flexible Electronics, Institute of Advanced Materials, and Jiangsu National Synergetic Innovation Center for Advanced Materials, Nanjing Tech University, Nanjing 211816, China

[†]These authors contributed equally to this work.
Email: sli@nju.edu.cn; yshi@nju.edu.cn

**ABSTRACT**: Antiferromagnetic (AF) materials are attracting increasing interest of research in magnetic physics and spintronics. Here, we report controllable synthesis of room-temperature AF α-MnTe nanocrystals (Néel temperature ~ 307 K) via molten-salt-assisted chemical vapor deposition method. The growth kinetics are investigated regarding the dependence of flake dimension and macroscopic shape on growth time and temperature. The high crystalline quality and atomic structure are confirmed by various crystallographic characterization means. Cryogenic magneto-transport measurements reveal anisotropic magnetoresistance (MR) response and a complicated dependence of MR on temperature, owing to the subtle competition among multiple scattering mechanisms of thermally excited magnetic disorders (magnon drag), magnetic transition and thermally populated lattice phonons. Overall positive MR behavior with twice transitions in magnitude is observed when out-of-plane external magnetic field ($B$) is applied, while a transition from negative to positive MR response is recorded when in-plane $B$ is applied. The rich magnetic transport properties render α-MnTe a promising material for exploiting functional components in magnetic devices.

## 1. INTRODUCTION

Antiferromagnetic (AF) materials have stimulated great interest in recent years because of various physical properties including electrical control of magnetism,[1,2] strong anomalous Hall effect,[3,4] and ultrahigh spin dynamics frequency,[5,6] which hold potential for exploiting memory devices,[7-9] novel magnetic detectors[10,11] and THz sources.[12,13] Compared with ferromagnets (FM), antiferromagnets possess antiparallel microscopic moments and zero net magnetization,[14,15] hence memories based on antiferromagnets would be robust against external magnetic disturbance, achieving high data retention and high-density integration.[16,17] In particular, AF semiconductors merging the magnetic order with electrical controlled transport properties of semiconductors have attracted widespread attention for their simultaneously conventional tunability of both charge and spin of carriers.[18,19] By taking advantages of electron spins as information carrier, such as non-volatile, free of Joule heating, long decoherence time and direct coupling to photon spins, the AF devices are expected to be faster, more functional and energy-efficient.[16,20,21]

Manganese telluride (MnTe) is a typical AF semiconductor with a moderate band gap of 1.25–1.46 eV.[22,23] Hexagonal NiAs-type α-MnTe is reported to be a stable phase under ambient conditions with a room-temperature Néel temperature ($T_N$) around 307 K.[24] Magnetic anisotropy,[25] susceptibility,[26] planar Hall effect,[27] thermal transport[28], and optical properties[29] of α-MnTe have been predicted or confirmed in experiments. The spintronic functionality of anisotropic magnetoresistance for electrical read-out of α-MnTe have also been exhibited, proving great potential in electronic variants of neural networks.[23] However, a systematic research on low-temperature magneto-transport properties of α-MnTe has been rare and the effect of the magnetic orders on the transport properties remains to be understood thoroughly before it could become an active element of real spintronic devices.[30] Besides, the growth techniques of α-MnTe in most researches are limited to molecular beam epitaxy, which is costly and time-consuming. As an important advance, Chang et al first reported the growth of α-MnTe via conventional chemical vapor deposition (CVD) method,[31] but the overall controllability in thickness and lateral size during synthesis is yet to be further improved, to meet the needs of miniaturized spintronic devices consisted of multiple layered functional layers.[32]

Herein, we report a simple and controllable synthesis strategy of α-MnTe crystalline flakes on mica substrates via the molten-salt-assisted CVD method. The growth kinetics were investigated with regard to the dependence of lateral dimension and flake shape on growth time and temperature. The high crystalline quality and atomic structure were characterized by X-ray diffraction (XRD) and high-resolution transmission electron microscopy (HRTEM). The unique magnetic structure-induced anisotropic magneto-transport behavior and temperature ($T$) dependent magnetoresistance response were uncovered, indicating a strong coupling of magnetic orders and electronic behavior in such an AF system, which is essential for magnetic physics and holds potential for exploiting novel spintronic devices.

## 2. RESULTS AND DISCUSSION

Figure 1a shows the top and side view of the atomic arrangement for the NiAs-type hexagonal α-MnTe crystal, which belongs to space group P63/mmc with the lattice parameters a = b = 4.158 Å, and c = 6.726 Å, described by the standard XRD card (PDF#18-0814). The unit cell of α-MnTe is marked with dotted line and also demonstrated in Figure 1b, where each Te atom is surrounded by six Mn atoms. Specifically, each Mn atom possesses a local spin moment oriented parallelly in the hexagonal basal plane but stacked antiparallelly along crystallographic c-axis.[25,27]

Hexagonal α-MnTe nanoflakes were grown by the molten-salt-assisted CVD method at atmospheric pressure, as illustrated in Figure 1c. Sodium chloride (NaCl) powders were finely grounded with manganese chloride ($MnCl_2$) to decrease the melting point of the reactants and increase the overall reaction rate. The similar mechanism was proved in the synthesis of a wide variety of transition-metal chalcogenides.[33] Besides, it is critical to choose an appropriate growth substrate to obtain the good controllability for target product. We first tried using conventional $SiO_2$/Si as substrates but failed to harvest thin and flat α-MnTe products, as shown in Figure S1 (Supporting Information). Generally, mica is a widely used substrate with a layered atomic structure and smooth dangling-bond-free surfaces, which allows precursors moving freely and is beneficial for horizontal growth of nonlayered material.[34] Figure 1d shows the α-MnTe flakes grown on mica, which are mainly triangularly shaped, conforming to the crystalline nature of α-MnTe. In addition, representative atomic force microscopy (AFM) characterization reveals the thickness of as-grown α-MnTe nanocrystals can be down to 13.5 nm, which is sufficiently thin for a non-layered material (Figure 1e and Figure S2, Supporting Information). More details about the growth process are provided in the Experimental Section.

To shed light on the atomic structure and crystalline quality of the as-grown α-MnTe flakes, XRD was employed to collect the diffraction patterns, as shown in Figure 1f, from which the diffraction patterns of our sample well matched the standard hexagonal α-MnTe data (PDF#18-0814). Noticeably, the XRD signals from the lattice planes other than the hexagonal basal plane, such as the (010), (101) and (012), are also detected, which are attributed to the presence of vertically grown α-MnTe flakes, as marked with red circles in Figure 1d and Figure S3 (Supporting Information). Besides, photoluminescence (PL) and Raman spectroscopies were used to characterize the magnitude of bandgap and the lattice vibration of the α-MnTe flakes. Figure 1g shows the presence of a strong lower optical absorption around 1.26 eV, whose position exhibits no notable dependence on thickness and shape. Raman spectra for a typical α-MnTe nanoplate and a mica substrate were shown in Figure S4 (Supporting Information) and strong vibration modes at 121 and 140 cm$^{-1}$ were recorded. Also, the elemental composition and related bonding states were examined by X-ray photoelectron spectroscopy (XPS). In Figure 1h, the two fitted peaks located at ~641.2 and ~653.2 eV correspond to the Mn $2p_{3/2}$ and Mn $2p_{1/2}$, respectively.[35,36] The other two peaks located at higher binding energies around 646.1 and 658.2 eV are satellites of the main peaks above.[36,37] In Figure 1i, the chemical states of Te $3d_{3/2}$ and $3d_{5/3}$ can be identified from the peaks at the binding energies of 572.4 and 582.9 eV.[38] We note that the other two peaks at positions of 575.9 and 586.3 eV likely derive from the oxidization of surface,[31,39] as tellurides are generally sensitive to air exposure,[40–43] which signifies the importance of surface protection after growth.

To further optimize the growth conditions, systematic experiments were conducted to reveal the effect of growth time and temperature on the lateral dimension and macroscopic shape of α-MnTe flakes (Figure 2). With the other growth parameters (temperature, source-substrate distance and carrier gas) unchanged, the α-MnTe nanosheets synthesized at various growth time exhibit notable differences in macroscopic flake size. Specifically, as presented in Figure 2a–d, the average flake size enlarges from 3 to 40 μm with extending the growth time from 5 to 30 min, in accompany with the augment of nucleation density and the evolution of flake shape. The dependence of flake size on growth time is summarized in Figure 2e. On the other hand, we also find the lateral dimension and macroscopic shape of the resultant flakes are quite sensitive to growth temperature. For instance, increasing the growth temperature from 500 to 540 °C, the averaged edge length of the triangular crystals increases from about 5 μm to about 15 μm. Whereas the flake shape evolves from perfect triangle, truncated triangle, hexagon and finally to round as the growth temperature changes from 500 to 600 °C, indicating a delicate thermal kinetics during growth. A summary of the evolution of lateral size and morphology of α-MnTe flakes with respect to growth temperature is exhibited in Figure 2f, where the increase of flake thickness can also be observed readily from the change of flake color under optical microscope.

To further understand the crystallographic structure of the CVD-grown α-MnTe flakes, TEM and selected-area electron diffraction (SAED) were performed on individual flakes. Here the as-grown α-MnTe flakes were transferred from mica substrates to copper TEM grids with carbon scaffolds via a poly(methyl-methacrylate) (PMMA) assisted transfer method. Figure 3a displays a typical low-magnification TEM image for a well-shaped trigonal α-MnTe flake on the TEM grids, manifesting the intactness of the flake after transfer. Figure 3b shows a typical atomically resolved crystallographic HRTEM image in which the atoms in hexagonal arrangement can be observed. The interplanar spacings of the two planes with 30° interfacial angle were measured to be 3.52 Å and 2.02 Å, respectively, corresponding to the (100) and (110) planes; the values of interplanar spacings also agree well with the standard XRD results (PDF#18-0814). The relevant SAED pattern (Figure 3c) indicates the high crystallinity of the CVD-grown α-MnTe flakes, as only one set of bright hexagonal diffraction spots are present. We further calculated the interplanar spacings of $d_{(110)}$ and $d_{(100)}$ basing on the SAED pattern, which were consistent with the values obtained from the XRD patterns. The chemical composition of a truncated triangular shaped α-MnTe flake was further detected by energy-dispersive spectroscopy (EDS, Figure 3d). It indicates that the as-grown flake well follows the stoichiometric ratio of 1:1 for Mn and Te elements. Elemental mapping images show that Mn and Te elements are uniformly distributed throughout the α-MnTe nanoflake (Figure 3e,f).

To gain insight into the rich magnetic physics in the low dimensional α-MnTe nanoflakes, cryogenic magneto-transport measurement was used to detect the response of electrical resistance to $T$ and $B$. It is worth noting that during device fabrication we minimized the exposure time of our samples in ambient as much as possible to avoid surface degradation and most procedures were performed in a nitrogen filled glove box, including sample inspection, subsequent transfer to the $SiO_2$/Si substrates and resist coating (Figure S5, Supporting Information). The transferred flakes were all metallized with standard Hall geometry on the $SiO_2$/Si substrates, as shown in the inset of Figure 4a (see Experimental Section for details). We first checked the dependence of the longitudinal resistivity ($\rho_{xx}$) of α-MnTe on temperature. Figure 4a illustrates a typical $\rho_{xx}$-$T$ curve for a 60 nm nanoflake from 320 to 2 K, which exhibits two evident transitions in magnitude around 307 and 120 K, respectively. The resistivity peak near 307 K is identified as the Néel temperature ($T_N$) of the α-MnTe, indicating the onset of AF phase below this temperature. The features of $\rho_{xx}$-$T$ were observed reproducibly in all devices measured (Figure S6, Supporting Information).

The transverse Hall resistance ($R_{xy}$) under varying magnetic fields was measured at various $T$s to monitor the density and type of charge carriers. As shown in Figure 4b, $R_{xy}$ exhibits a positive and rather linear correlation with $B$, implying that the holes are the majority charge carriers and a simple one-band model can be used to estimate the carrier density. Figure 4c summarizes the values of estimated Hall carrier density at different $T$s. It changes monotonically from $4.7 \times 10^{13}$ cm$^{-2}$ at 20 K to $1.1 \times 10^{14}$ cm$^{-2}$ at 150 K and then becomes nearly saturated up to 280 K. The magnitude of carrier density measured here is consistent with the values of α-MnTe reported in literature,[23] and its trend also coincides with the $\rho_{xx} - T$ curve. In detail, the drop of $\rho_{xx}$ from 2 to 120 K arises presumably from the increase of hole concentration at elevated $T$, while the rise of $\rho_{xx}$ above 120 K can be attributed to enhanced scatterings from phonons and magnons.[44] Specifically, the weak linear increase of $\rho_{xx}$ from 120 to 150 K reflects the phonon contribution while the strong rise of $\rho_{xx}$ above the linear phonon part (from 150 K to $T_N$) originates from the thermally excited magnons,[26,28,44,45] i.e., interaction of the itinerant electrons with localized spins, where the phonon contribution is much smaller.

We then performed MR measurements at various $T$s for both conditions: $B \parallel$ c axis and $B \parallel$ ab plane of the α-MnTe nanoflakes, as shown in Figure 4d–f. Here MR is defined as $\mathrm{MR} = \frac{\rho_{xx}(B) - \rho_{xx}(0)}{\rho_{xx}(0)} \times 100\%$, denoting the change in the percentage of longitudinal resistivity under magnetic fields. In case of $B \parallel$ c axis, as shown in Figure 4d,e, MR is positive and non-saturated under all applied $B$ and $T$ conditions. As expected, MR is quadratically dependent on $B$ within 3 T, in line with the Kohler's rule. The $T$-dependent MR of the α-MnTe is further summarized in Figure 4g where we see a non-monotonic MR under $B$ = 1, 2 and 3 T. For instance, MR firstly increases with $T$ by varying from ~0.6% at 10 K to ~1.5% at 120 K, followed by an obvious drop to ~0.5% at 300 K, and finally rises again to ~0.9% at 320 K under 3 T. These characteristic $T$ nodes are totally consistent with those in the $\rho_{xx}$-$T$ curve, as a consequence of the subtle competition between multiple thermally associated scattering (i.e., lattice phonons and magnons) and phase transition mechanisms. In detail, the drop of MR below 120 K arises likely from the sharp rise in susceptibility, due to magneto-elastic coupling that enhances intraplanar ferromagnetic interactions relative to interplanar AF interactions caused by the increase of lattice parameters c/a ratio.[26,46] In other words, the increase of magnetic ordering can effectively reduce the spin-dominant scattering component at $T$ < 120 K. Afterwards, the drop of MR from 120 to 300 K reflects the enhanced effect of magnon drag.[47,48] The turning point of MR around 300 K suggests the occurrence of AF-paramagnetic (PM) transition, and the rise of MR above 300 K could be attributed to field-induced PM magnetization.[49]

Figure 4f shows the MR characteristics as $B \parallel$ ab plane in which a reverse of sign is observed near $T_N$. The negative MR below $T_N$ originates likely from the suppressed scattering of local spin fluctuations by increasing long-range magnetic ordering when $B$ is parallel to the spin moments ordered Mn planes.[20,50,51] This feature confirms the magnetic structure of our CVD-grown α-MnTe, i.e. the magnetic moments are oriented in the ab plane, as shown in Figure 1b. The negative MR remains unsaturated under $B$ = 3 T with the magnitude of ~−0.7% at 180 K. Upon increasing $T$, similarly, the in-plane MR becomes increasingly weak and finally disappears around $T_N$ (Figure 4h), which can be attributed to the increase of thermally populated magnons and the enhanced dominance of magnon drag effect at elevated $T$ regime.[47,48] Finally, the sign of the in-plane MR is reversed to positive after the AF-PM transition.[52]

## 3. CONCLUSION

In summary, high quality antiferromagnetic semiconductor α-MnTe crystalline flakes were controllably synthesized on mica substrates via molten-salt-assisted CVD method. Various means including XRD, SAED and HRTEM were employed to characterize and check the atomic structure and crystallinity of the as-synthesized samples. The AF-PM transition and the rivalry between thermally excited magnons and lattice phonons were uncovered by $T$-dependent $\rho_{xx}$ and MR. In particular, the reverse of sign of MR around $T_N$ is observed when $B$ is applied along the in-plane direction. The room-temperature $T_N$ and strong coupling of magnetic characteristics to electronic behavior make α-MnTe attractive to construct $T$ and $B$ sensitive functional components in spintronics.

## 4. EXPERIMENTAL SECTION

**Synthesis of Hexagonal α-MnTe Flakes:** High-quality α-MnTe nanocrystals were grown in a multitemperature-zone tube furnace equipped with a 34 mm diameter quartz under ambient pressure. Te powder (Aladdin, 98%) (0.1 g) was placed in the first hot zone, rapidly heating up to 600 °C later. A mixture of well grounded $MnCl_2$ (Aladdin, 99%) and NaCl (Aladdin, 99.5%) (30 mg) with the weight ratio of 3:1 was placed at the center of next heating zone. A piece of freshly cleaved fluorophlogopite ($[KMg_3(AlSi_3O_{10})F_2]$) mica sheet (Nanjing MKNANO Tech.) was located at the downstream about 1 cm away from the $MnCl_2$/NaCl source powder. In view of the fact that $MnCl_2$ is sensitive to water vapors in the air, the CVD system needs to be preheated to about 150 °C and evacuated for a dry and anaerobic environment. After that, the system was purged with Ar gas to restore ambient pressure. Then the furnace temperature was ramped up to 500 °C within 21 min and held for 15 min under a constant flow of Ar (200 sccm) and $H_2$ gas (50 sccm). Finally, the furnace was cooled down to room temperature.

**Transfer of CVD-Grown α-MnTe Flakes:** The as-grown α-MnTe flakes on mica can be transferred onto an arbitrary substrate with a wet transfer approach. In brief, the mica substrate was spin-coated with PMMA 950K A4 solution at 1000 rpm for 60 s and baked for 3 min at 180 °C. The operation was then repeated once to ensure a sufficient thickness of PMMA to prevent tear-up during transfer. Then the α-MnTe flakes covered by PMMA were detached from mica by tweezers after immersion in deionized water for 1.5 h and put onto a target substrate (e.g., $SiO_2$/Si). At last, the transferred specimen was dried on a heater to remove the water between PMMA and substrate, followed by acetone immersion to dissolve the PMMA. Samples for TEM characterization were prepared by the same way.

**Device Fabrication and Transport Measurement:** The as-grown α-MnTe flakes were quickly moved to a nitrogen-filled glove box for transfer onto $SiO_2$/Si substrates and resist coating. The electron resist PMMA 950K A4 was used and the recipe for resist coating is 4000 rpm, 1 min. The Hall bar devices were fabricated by standard e-beam lithography, thermal evaporation of 5 nm Ni/100 nm Au bilayers as electrodes and standard lift-off. All electrical transport experiments were performed in a cryostat with $T$ from 320 to 1.5 K and $B$ up to 12 T (Oxford Instruments). The longitudinal and Hall voltages were measured by lock-in amplifiers (Sine Scientific Instruments).

**Characterization:** The facilities for material characterization include optical microscopy (Olympus BX53M), Raman spectrometer (WITEC alpha 300R confocal micro-Raman system using a 532 nm laser as the excitation source), AFM (Asylum Research Cypher system with tapping mode in ambient), XRD (Bruker Discover8 with Cu-Kα emission, wavelength=1.5418 Å), XPS (Thermo Scientific Escalab 250Xi) and TEM (Tecnai F20, FEI) equipped with an EDS.

## Supporting Information

Optical images of flakes grown on $SiO_2$/Si, AFM image for the thinnest sample observed, optical images of vertical growth, Raman spectra of α-MnTe flake, photograph of the glove box and longitudinal resistivity versus temperature of α-MnTe.

**Acknowledgments**

The work was supported by the National Key R&D Program of China (2017YFA0206304), the National Natural Science Foundation of China (61974060, 61674080, 61521001 and 11874199), the Innovation and Entrepreneurship Program of Jiangsu province and the Micro Fabrication and Integration Technology Center in Nanjing University.

**Figure captions**

**Figure 1.** Molten-salt-assisted CVD synthesis of α-MnTe crystals and characterizations. (a) Top and side view of the atomic arrangement. The pink and grey balls represent the Mn and Te atoms, respectively. (b) Unit cell with local spin moments exhibited on Mn atoms. (c) Schematic illustration for the CVD setup. (d) Typical optical image for α-MnTe flakes. The vertically grown flakes are marked with red circles. Scale bar: 20 μm. (e) AFM surface topography and corresponding height profile for a triangular flake. Scale bar: 5 μm (f) XRD pattern. (g) PL spectra for flakes with varied shapes and thicknesses. Inset: Corresponding AFM images. Scale bar: 5 μm. XPS characterization for (h) Mn 2p and (i) Te 3d peaks.

**Figure 2.** Evolution of the dimension and macroscopic shape of α-MnTe nanoflakes under different synthesis conditions. Typical optical images for α-MnTe nanoplates grown at increasing growth time of (a) 5 min, (b) 15 min, (c) 20 min, (d) 30 min, respectively. Scale bar: 50 μm. (e) Dependence of lateral size on growth time. (f) Evolution of lateral size and shape at varied growth temperatures from 500 to 600 °C.

**Figure 3.** TEM characterization of as-grown α-MnTe crystals. TEM images for a triangular α-MnTe flake at (a) low magnification and (b) high resolution. (c) Corresponding electron diffraction pattern. (d) EDS spectrum for another MnTe flake. Inset: Corresponding TEM image of the selected area. Elemental mapping images for the elements (e) Mn and (f) Te, respectively.

**Figure 4.** Cryogenic magneto-transport properties of α-MnTe nanoflakes. (a) $T$-dependent resistivity. Inset: Optical image of the device in Hall geometry. Scale bar: 5 μm. (b) Hall resistance at different $T$s. (c) Extracted carrier density as a function of $T$ with one-band model. $T$-varied MR with $B$ parallel to (d,e) c axis and (f) ab plane, respectively. Summaries of the dependence of MR on $T$ at various $B$ directions: (g) out-of-plane and (h) in-plane.

# Figures

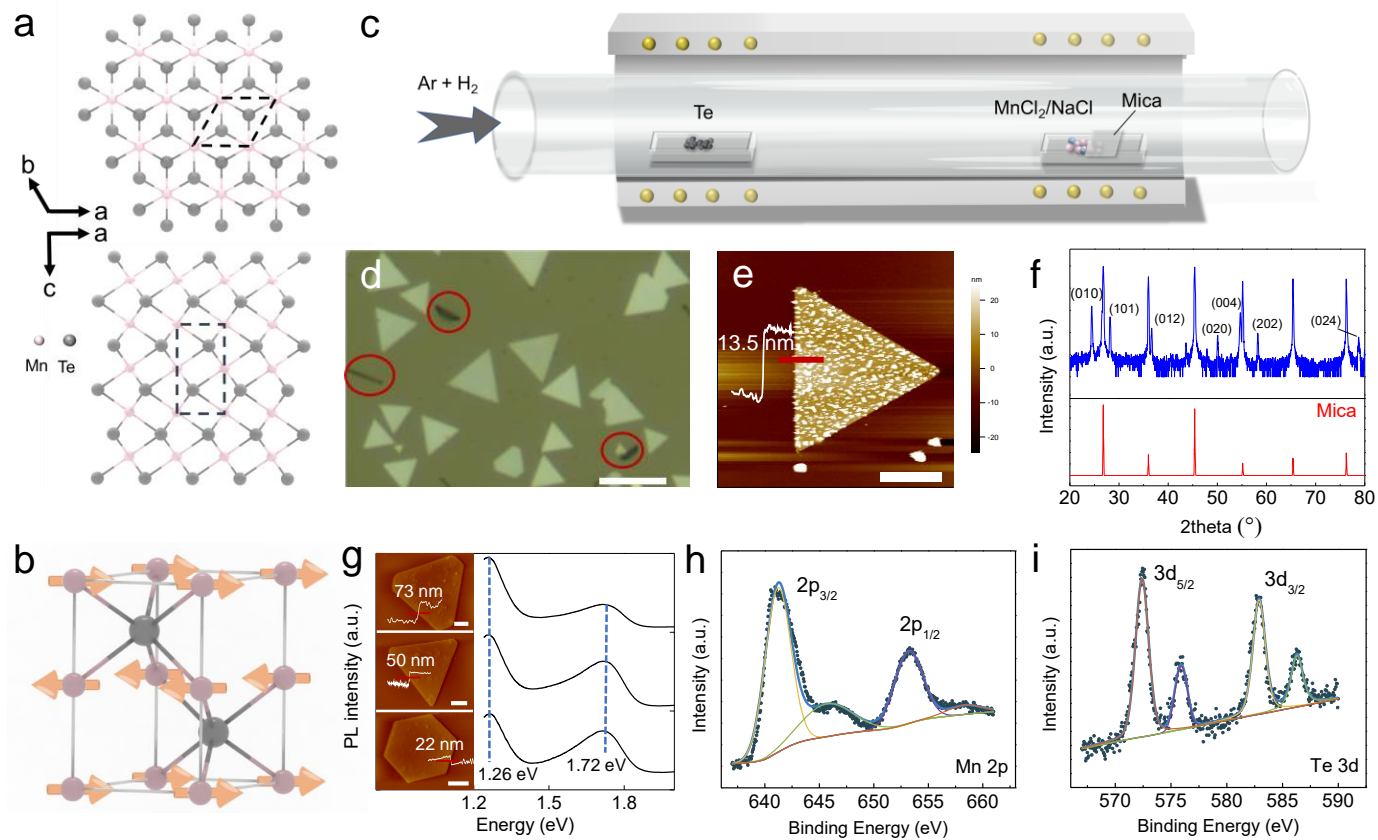

Figure 1.

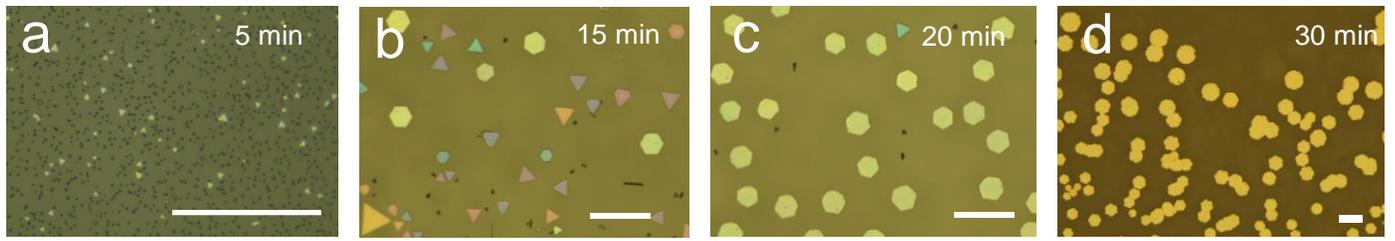
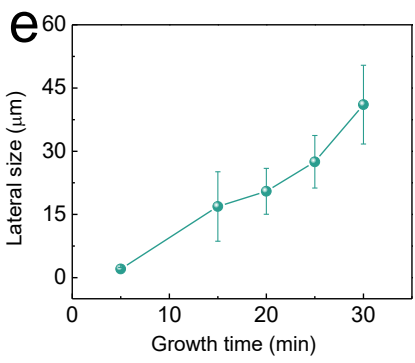
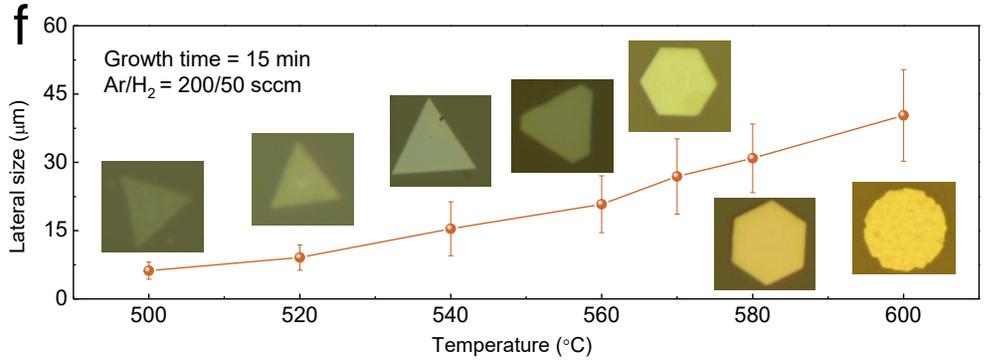

Figure 2.

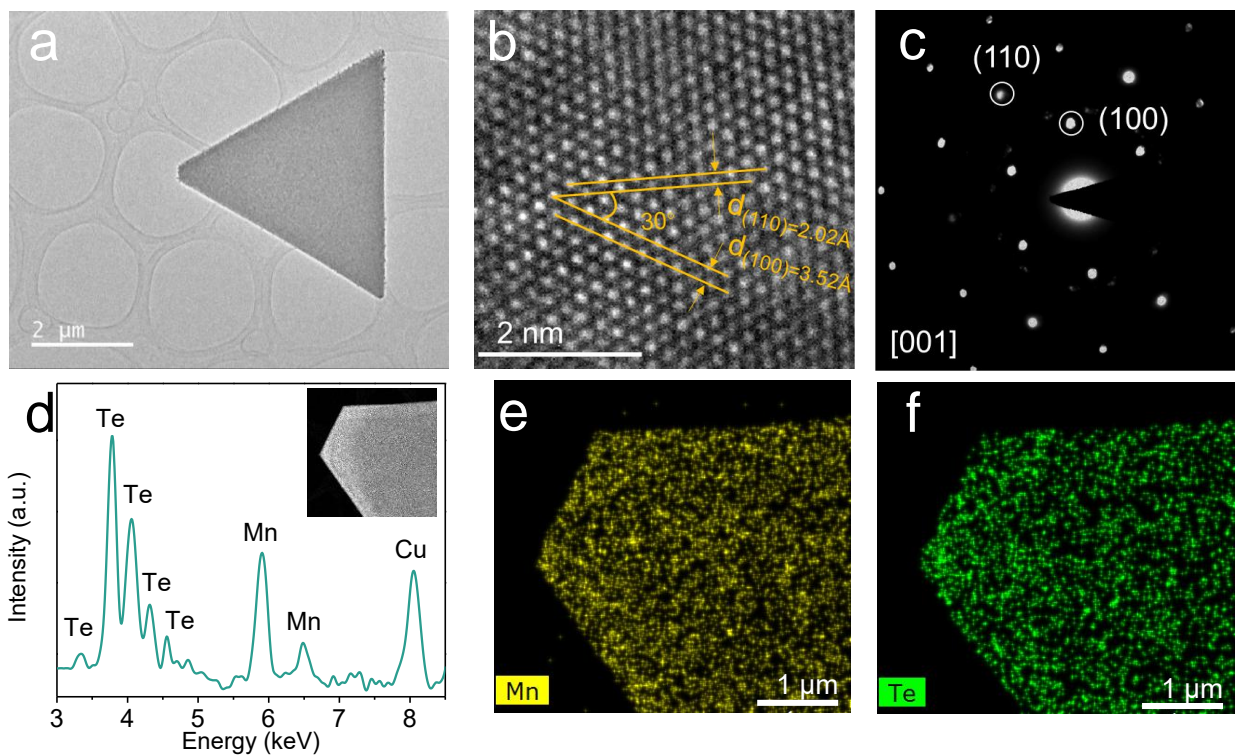

Figure 3.

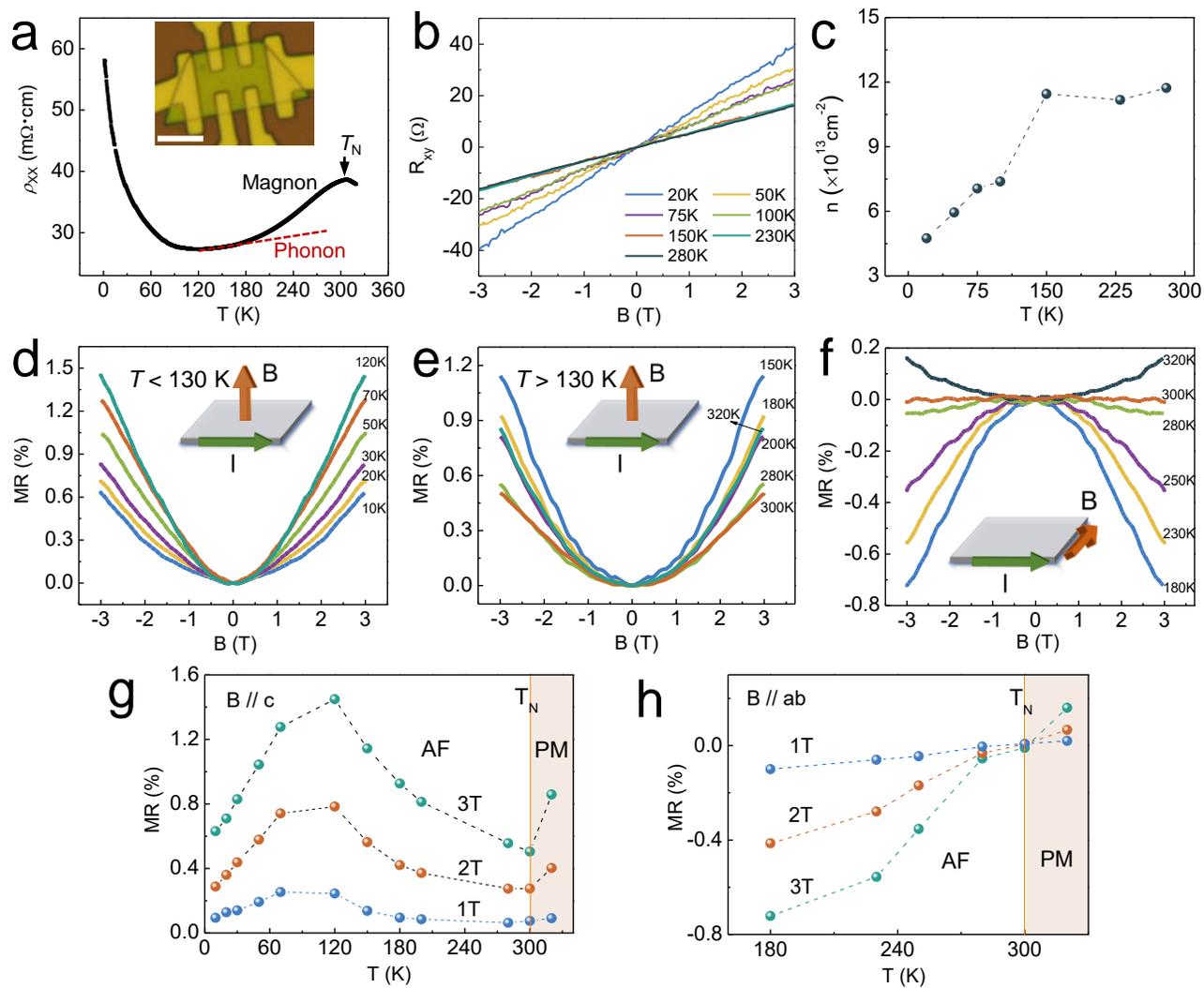

Figure 4.